\documentclass[superscriptaddress,
prx,
preprintnumbers,
amsmath,
amssymb,
altaffilletter,
floatfix,
]{revtex4-2}

\usepackage{graphicx}
\usepackage{bm}
\usepackage{caption}
\usepackage{wrapfig}

\usepackage{subcaption}
\usepackage{url}
\usepackage{orcidlink}

\begin{document}

\title{EXO-200 Public Data Release for AI/ML Applications}

\newcommand{\instStanfordPhys}{Physics Department, Stanford University, Stanford, California 94305, USA}
\newcommand{\instMcGillPhys}{Physics Department, McGill University, Montreal, Quebec H3A 2T8, Canada}
\newcommand{\instKACST}{King Abdulaziz City for Science and Technology, Riyadh, Saudi Arabia}
\newcommand{\instCarletonPhys}{Physics Department, Carleton University, Ottawa, Ontario K1S 5B6, Canada}
\newcommand{\instDukeTUNL}{Department of Physics, Duke University, and Triangle Universities Nuclear Laboratory (TUNL), Durham, North Carolina 27708, USA}
\newcommand{\instIllinoisPhys}{Physics Department, University of Illinois, Urbana-Champaign, Illinois 61801, USA}
\newcommand{\instITEPKurchatov}{Institute for Theoretical and Experimental Physics named by A.I. Alikhanov of National Research Centre ``Kurchatov Institute,'' Moscow 117218, Russia \thanks{Now a division of National Research Center ``Kurchatov Institute,'' Moscow 123182, Russia}}
\newcommand{\instUSDakotaPhys}{Department of Physics, University of South Dakota, Vermillion, South Dakota 57069, USA}
\newcommand{\instUKentuckyPhys}{Department of Physics and Astronomy, University of Kentucky, Lexington, Kentucky 40506, USA}
\newcommand{\instSLAC}{SLAC National Accelerator Laboratory, Menlo Park, California 94025, USA}
\newcommand{\instTRIUMF}{TRIUMF, Vancouver, British Columbia V6T 2A3, Canada}
\newcommand{\instIHEP}{Institute of High Energy Physics, Beijing 100049, China}
\newcommand{\instWitmem}{Witmem Technology Co., Ltd., No.56 Beisihuan West Road, Beijing, China}
\newcommand{\instLaurentianPhys}{Department of Physics, Laurentian University, Sudbury, Ontario P3E 2C6, Canada}
\newcommand{\instCSUPhys}{Physics Department, Colorado State University, Fort Collins, Colorado 80523, USA}
\newcommand{\instUNCWPhys}{Department of Physics and Physical Oceanography, University of North Carolina at Wilmington, Wilmington, NC 28403, USA}
\newcommand{\instIndianaCEEM}{Physics Department and CEEM, Indiana University, Bloomington, Indiana 47405, USA}
\newcommand{\instDrexelPhys}{Department of Physics, Drexel University, Philadelphia, Pennsylvania 19104, USA}
\newcommand{\instECAPFAU}{Erlangen Centre for Astroparticle Physics (ECAP), Friedrich-Alexander-University Erlangen-N\"urnberg, Erlangen 91058, Germany}
\newcommand{\instTUMUniverse}{Technische Universit\"at M\"unchen, Physikdepartment and Excellence Cluster Universe, Garching 80805, Germany}
\newcommand{\instUVAPhys}{Department of Physics, University of Virginia, Charlottesville, VA 22904, USA}
\newcommand{\instUmass}{Amherst Center for Fundamental Interactions and Physics Department, University of Massachusetts, Amherst, MA 01003, USA}
\newcommand{\instUMDPhys}{Physics Department, University of Maryland, College Park, Maryland 20742, USA}
\newcommand{\instUCBerkeleyPhys}{Department of Physics at the University of California, Berkeley, California 94720, USA}
\newcommand{\instYaleWright}{Wright Laboratory, Department of Physics, Yale University, New Haven, Connecticut 06511, USA}
\newcommand{\instPrincetonPhys}{Department of Physics, Princeton University, Princeton, New Jersey, USA}
\newcommand{\instIBSCUP}{IBS Center for Underground Physics, Daejeon 34126, Korea}
\newcommand{\instSBUPhys}{Department of Physics and Astronomy, Stony Brook University, SUNY, Stony Brook, New York 11794, USA}
\newcommand{\instAlabamaPhys}{Department of Physics and Astronomy, University of Alabama, Tuscaloosa, Alabama 35487, USA}
\newcommand{\instCanonMedical}{Canon Medical Research US Inc., Vernon Hills, IL, USA}
\newcommand{\instIowaStatePhys}{Department of Physics and Astronomy, Iowa State University, Ames, IA 50011, USA}
\newcommand{\instCaltechKellogg}{Kellogg Lab, Caltech, Pasadena, California 91125, USA}
\newcommand{\instBernLHEP}{LHEP, Albert Einstein Center, University of Bern, Bern, Switzerland}
\newcommand{\instDescartesLabs}{Descartes Labs, 100 North Guadalupe, Santa Fe, New Mexico 87501, USA}
\newcommand{\instHamburgIEP}{Institute for Experimental Physics, Hamburg University, 22761 Hamburg, Germany}
\newcommand{\instSNOLAB}{SNOLAB, Sudbury, ON, Canada}
\newcommand{\instLBNL}{Lawrence Berkeley National Laboratory, Berkeley, California, USA}
\newcommand{\instFermilab}{Fermilab, Batavia, IL 60510, USA}
\newcommand{\instSCIPP}{SCIPP, University of California, Santa Cruz, California, USA}
\newcommand{\instUCSDPhys}{Physics Department, University of California, San Diego, La Jolla, California 92093, USA}
\newcommand{\Hawaii}{Department of Physics and Astronomy, University of Hawaii at Manoa, Honolulu, HI 96822, USA}
\newcommand{\windsor}{Department of Physics, University of Windsor, Windsor, ON N9B 3P4, Canada}
\newcommand{\addrLosAngeles}{Los Angeles, California 90025, USA}

\author{S.~Al~Kharusi}
  \altaffiliation{Present address: \instStanfordPhys}
  \affiliation{\instMcGillPhys}

\author{G.~Anton}
  \affiliation{\instECAPFAU}

\author{I.~Badhrees}
  \altaffiliation{Permanent address: \instKACST}
  \affiliation{\instCarletonPhys}

\author{P.S.~Barbeau}
  \affiliation{\instDukeTUNL}


\author{V.~Belov}
  \affiliation{\instITEPKurchatov}

\author{T.~Bhatta}
  \altaffiliation{Present address: \instUKentuckyPhys}
  \affiliation{\instUSDakotaPhys}

\author{M.~Breidenbach}
  \affiliation{\instSLAC}

\author{T.~Brunner}
  \affiliation{\instMcGillPhys}
  \affiliation{\instTRIUMF}

\author{G.F.~Cao}
  \affiliation{\instIHEP}

\author{W.R.~Cen}
  \altaffiliation{Present address: \instWitmem}
  \affiliation{\instIHEP}

% ...

\date{September 27, 2026}

\begin{abstract}
We present a public release of a subset of calibration data from the EXO-200 experiment, a liquid xenon time projection chamber designed to search for neutrinoless double-beta decay in $^{136}$Xe. The released dataset consists of approximately one million events collected during $^{228}$Th calibration campaigns performed near the end of Phase-II operations between February 26 and March 2, 2018. For each event, the dataset includes raw detector waveforms from the U-wire, V-wire, and avalanche photodiode (APD) readout channels together with reconstructed event quantities including charge energy, light energy, rotated energy, and charge-cluster information. The dataset is distributed in HDF5 format and is intended to support data preservation, educational activities, and the development of modern machine-learning techniques for rare-event physics. This paper describes the detector, dataset contents, file structure, and public access mechanism.
\end{abstract}

\maketitle


\section{Introduction}

\vspace{-5pt}

The rapid development of artificial intelligence (AI) and machine learning (ML) has increased demand for large, high-quality scientific datasets. At the same time, federal agencies and scientific organizations have emphasized the importance of open science and public access to research data~\cite{nelson}. Publicly available datasets improve reproducibility, broaden participation, and often enable new scientific applications beyond the original goals of an experiment.

Particle and astroparticle physics experiments produce uniquely valuable datasets consisting of high-dimensional detector signals accompanied by carefully validated reconstruction outputs. Recent public releases from experiments including MicroBooNE~\cite{cerati2023microboonepublicdatasets}, IceCube~\cite{Bukhari2024IceCube}, COHERENT~\cite{COHERENT}, and the Majorana Demonstrator~\cite{MJD-datarelease} have demonstrated the scientific and educational value of making detector data broadly accessible.


The EXO-200 experiment completed data taking in 2018. In support of open science and data preservation, we release a subset of EXO-200 calibration data designed to provide a representative sample of liquid xenon detector signals. The released dataset contains waveform-level detector information together with reconstructed quantities commonly used in EXO-200 analyses. The release is intended to serve as a community resource for detector studies, algorithm development, educational activities, and machine-learning research.

This document is structured as follows: Section~\ref{sec:detector} briefly discusses the EXO-200 detector. Section~\ref{sec:content} provides an overview of the dataset's content, format, and methods for accessing it;  Section~\ref{sec:disclaimer} contains a disclaimer from the EXO-200 collaboration regarding the use of this dataset.

\begin{wrapfigure}{r}{0.35\textwidth}
\vspace{-10pt}
\centering
\includegraphics[width=0.37\textwidth,trim=50 0 70 0, clip]{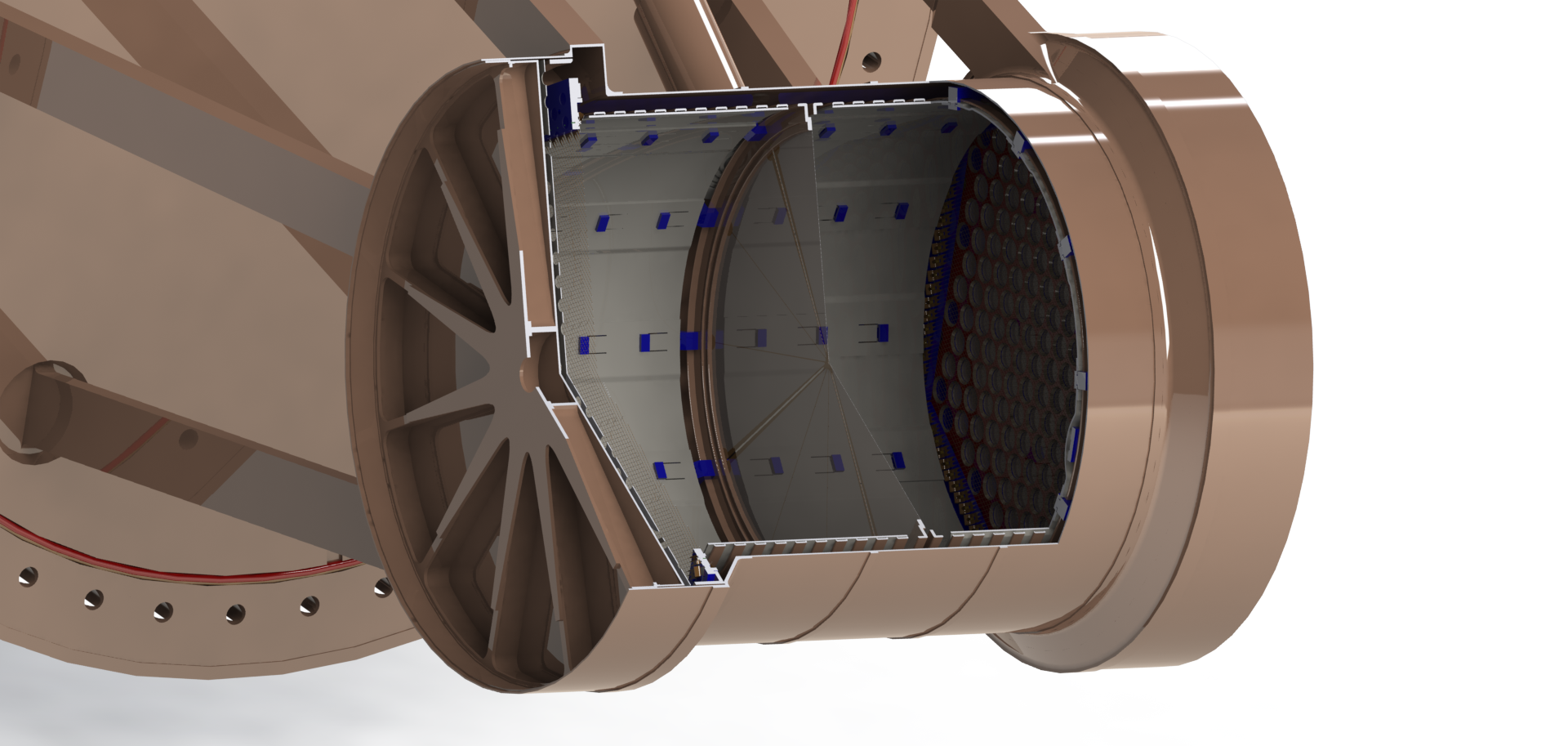}
\caption{Cutaway view of the EXO-200 TPC.}
\label{fig:exo_tpc}
\vspace{-10pt}
\end{wrapfigure}

\vspace{-10pt}

\section{The EXO-200 Detector}
\label{sec:detector}

\vspace{-5pt}
EXO-200 was the first double-beta-decay experiment at the 100-kg scale and the first to observe the standard $2\nu\beta\beta$ decay mode of $^{136}$Xe~\cite{Ackerman_2011}. The experiment was located at the Waste Isolation Pilot Plant (WIPP) near Carlsbad, New Mexico, and set some of the most stringent limits on the $0\nu\beta\beta$ decay mode of $^{136}$Xe~\cite{EXO-0nbb-12,EXO-200-onbb-14, Albert_2018, Anton_2019}. 
The detector contained liquid xenon (LXe) enriched to approximately 80\% in $^{136}$Xe, serving as both the source material and the detector medium. 
The cylindrically symmetric outer vessel is made of ultra-low activity copper.
The central component of the detector is a pair of LXe time projection chambers (TPCs).     
The detector is divided into two equal-volume TPCs by a shared cathode plane.  
At the end of each detector is a pair of wire planes crossed at 60 degrees that form the anodes.
Behind the wires is an array of large-area avalanche photodiodes (LAAPDs)~\cite{Neilson_2009} looking into the TPC active 
\clearpage{}
\noindent
volumes.  
Field-shaping rings 
surround the TPC volume to improve
drift field uniformity.
The TPC barrels are lined with Teflon reflectors to improve light-collection efficiency. Of EXO-200's $200~\rm{kg}$ xenon inventory, 
 $\sim\!175~\rm{kg}$ is liquefied, and $\sim\!110~\rm{kg}$ is contained within the TPCs. 
When an ionizing radiation event occurs inside a TPC, the ionization electrons
are drifted towards the wire planes by a main drift field applied between the wires and the cathode. The ``V-wire" plane, encountered first by the drifting electrons, records induction signals, while the ``U-wire" plane collects the charge. The wire signals provide two-dimensional position information of the event.  The position along the drift direction is reconstructed from the time difference between the prompt scintillation signal and the charge arrival time, using the known electron drift speed. Both the wire and APD signals are read out by charge-sensitive preamplifiers outside the lead shielding. 
The front-end electronics shape the signals using three differentiation and two integration stages, with time constants optimized for each signal type; details on the shaping times can be found in \cite{shaolei_thesis}.
More details about the detector can be found in Refs.~\cite{EXO-Detector_2012, EXO-detector_2022, Albert_2014}. 

\section{Dataset description}\label{sec:content}

\subsection{Dataset Content}
\begin{figure*}[h]
\centering

\begin{minipage}[t]{0.50\textwidth}
    \centering
    \includegraphics[width=\linewidth]{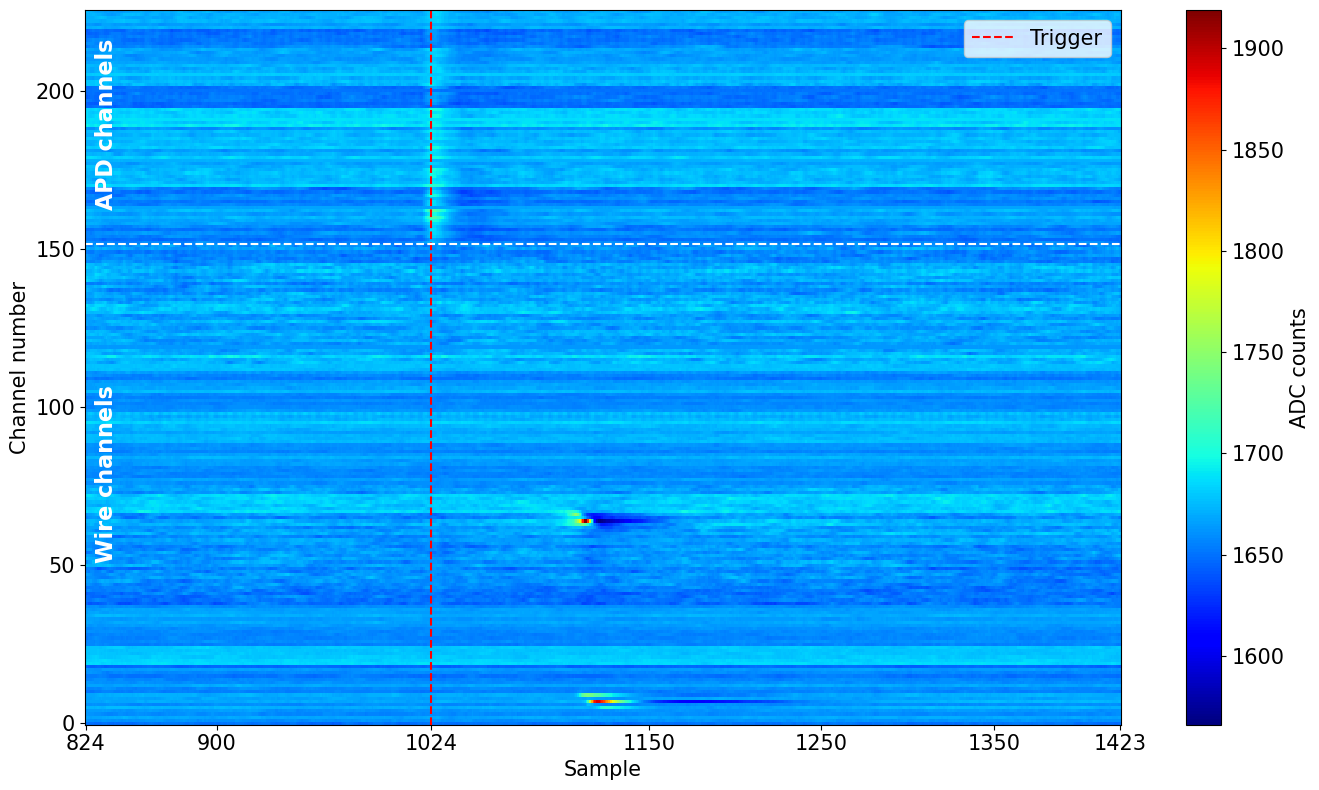}
    \vspace{0.3em}
 
\end{minipage}
\hfill
\begin{minipage}[t]{0.48\textwidth}
    \centering
    \includegraphics[width=\linewidth]{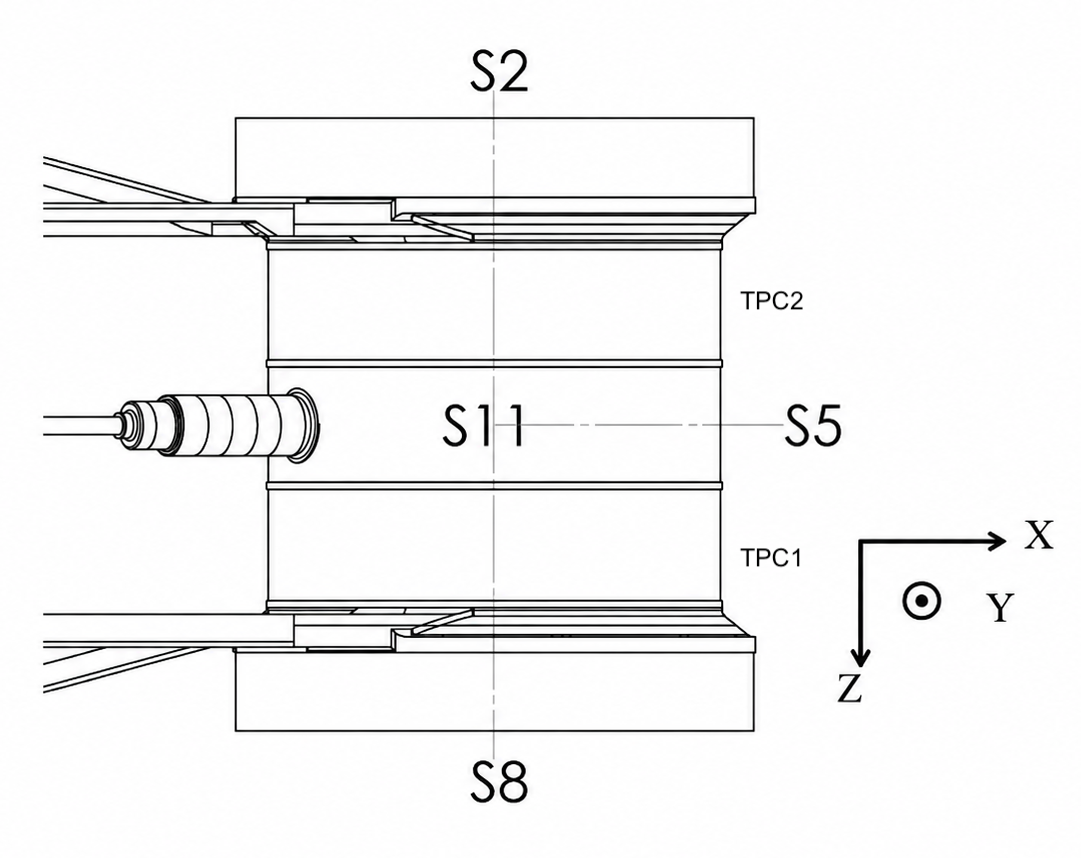}
    \vspace{0.3em}

\end{minipage}

\caption{Left: Example EXO-200 raw waveforms including 76 U-wire channels, 76 V-wire channels, and 74 APD channels. The trigger at $1024~\mu\rm{s}$ in this event is on the summed APD signal. Right: Schematic illustration of the $^{228}$Th source deployment locations used during the calibration campaign.
}
\label{fig:waveform_source}
\end{figure*}

The EXO-200 public dataset includes both raw waveform data and high-level reconstructed quantities used in the EXO-200 physics analyses. 
The dataset consists of $^{228}$Th calibration runs collected during the quarterly calibration campaign near the end of Phase-II operations, spanning the period from 2018-02-26 to 2018-03-02. 
These runs were acquired with the source at four separate source tube positions: S2, S5, S8, and S11. See Figure~\ref{fig:waveform_source} (Right) for the relative locations of the source positions in TPC coordinates. 
Information for the calibration runs is summarized in Table~\ref{tab:exoruns}.  
The 15 selected runs contain approximately 2.2 million triggered events before cuts. We apply an energy cut requiring the charge, scintillation, and rotated energies of each event to fall within 500–4,000 keV. This cut removes approximately 52.3\% of the events. 
Additional cuts limit the number of reconstructed charge clusters to a maximum of five and require the largest charge cluster to be successfully reconstructed within the TPC; these remove a further 2.6\% of the events. The entire public dataset thus consists of $\sim$ 1 million events.


\vspace{1em}

\begin{table}[hbt]
\centering
\caption{Summary of the EXO-200 calibration public dataset. All runs are of Thorium source calibration taken during the calibration campaign, 2018-02-26 to 2018-03-02.}
\begin{tabular}{c|c|c|c}
\hline
Source Position & Run Numbers & Number of Runs & Total Exposure (hrs) \\
\hline
S2 & 9008--9010 & 3 & 6 \\
S5 & 8967--8972 & 6 & 11 \\
S8 & 8956--8958 & 3 & 6 \\
S11 & 8999--9001 & 3 & 6 \\
\hline
Total &  & 15 & 29 \\
\hline
\end{tabular}
\label{tab:exoruns}
\end{table}

\begin{figure}
    \centering
    \includegraphics[width=0.9\linewidth]{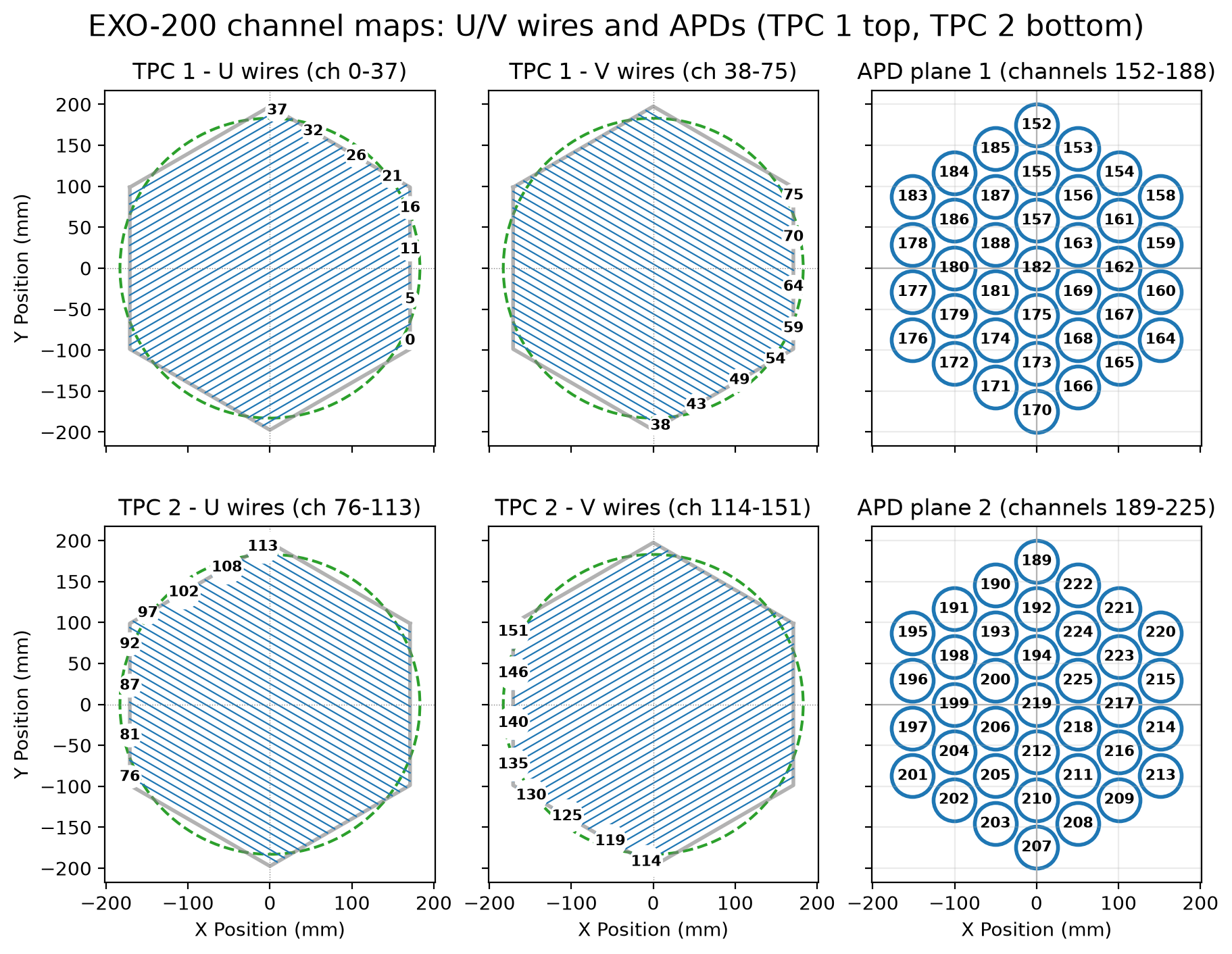}
    \caption{The locations of each channel in TPC coordinates for the U-Wires, V-wires, and APD gangs. The hexagon frame shows the extent of the wires, and the green circle shows the extent of the Teflon reflector. APD channels 163, 178, 191, and 200 are disconnected due to noise issues. Channel 205 is connected to channel 202 during detector construction.}
    \label{fig:channel_map}
\end{figure}

Each event within the public dataset includes the quantities summarized in Table~\ref{tab:exodescription}, including truncated U/V-wire and APD waveforms spanning a 600~$\mu$s window around the hardware trigger time (200~$\mu$s before and 400~$\mu$s after the trigger). Figure~\ref{fig:waveform_source} (left) shows an example of the truncated waveform in the dataset. A hardware trigger can occur in several scenarios, but the most common is that the sum of all APD signals is above threshold, and the second most common is a U-wire signal above threshold. This means that the hardware trigger can occur on either the light or charge signals. The channel locations in the waveform array are 
shown in Figure~\ref{fig:channel_map} in the TPC coordinate space. 

The dataset also includes reconstructed event quantities, including charge energy, light energy, rotated energy (a linear combination of charge and light energies that optimizes the detector resolution), charge cluster energy information for up to 5 reconstructed clusters, and the position of the largest charge cluster. The charge energy is reconstructed using the EXO-200 signal-fitting and clustering algorithms and corrected for electronic gain, grid-shielding effects, and electron attenuation due to finite LXe purity, but it is not independently calibrated using the calibration-source spectra.~\cite{Albert_2014}. The scintillation energy is reconstructed using a denoising algorithm that accounts for temporal and spatial variations in the light response while minimizing the impact of electronic noise~\cite{Davis_2016}. The rotated energy is fully calibrated using the complete set of calibration-source data and provides the best estimator of the event energy~\cite{Anton_2019}. In addition, the measured electron lifetime for the calibration period is included in the open-source dataset. More details on EXO-200 data analysis strategy and reconstruction procedures can be found in Ref.~\cite{Albert_2014, Anton_2019, Delaquis2018}.

\renewcommand{\arraystretch}{1.05}

\begin{table*}[t]
\centering
\small

\caption{Description of each event in the EXO-200 open-source dataset.}
\label{tab:exodescription}
\begin{tabular}{p{5.7cm}|p{2.5cm}|p{9.0cm}}
\hline
 Dataset Field Name & Data Type & Description \\
\hline

\texttt{Event\_Number} &
Integer &
Assigned unique identifier for each event in a run
\\
\hline

 \texttt{Waveforms} &
Two-dimensional Integer Array &
Raw waveforms from U, V wire and APD channels, truncated to 600 $\mu$s near the trigger time. (raw ADC data) 
\\
\hline

\texttt{Charge\_Cluster\_Number}  &
Integer &
Number of reconstructed charge clusters in the event
\\
\hline

 \texttt{Charge\_Cluster\_Energy} &
Variable-length Float Array &
Reconstructed charge energy in keV for up to 5 clusters.
\\
\hline

\texttt{Charge\_Cluster\_Position} &
Float Array &
Reconstructed position (x, y, z) of the largest charge cluster
\\
\hline

 \texttt{Total\_Charge\_Cluster\_Energy} &
Float &
Total reconstructed charge energy of the event in keV
\\
\hline

\texttt{Total\_Scintillation\_Cluster\_Energy} &
Float &
Total reconstructed scintillation energy of the event in keV
\\
\hline

 \texttt{Rotated\_Energy} &
Float &
Rotated energy of the event in keV
\\
\hline

\end{tabular}

\end{table*}










\subsection{Dataset access}\label{sec:access}

\begin{figure*}[h]
\centering

\begin{minipage}[t]{0.48\textwidth}
    \centering
    \includegraphics[width=\linewidth]{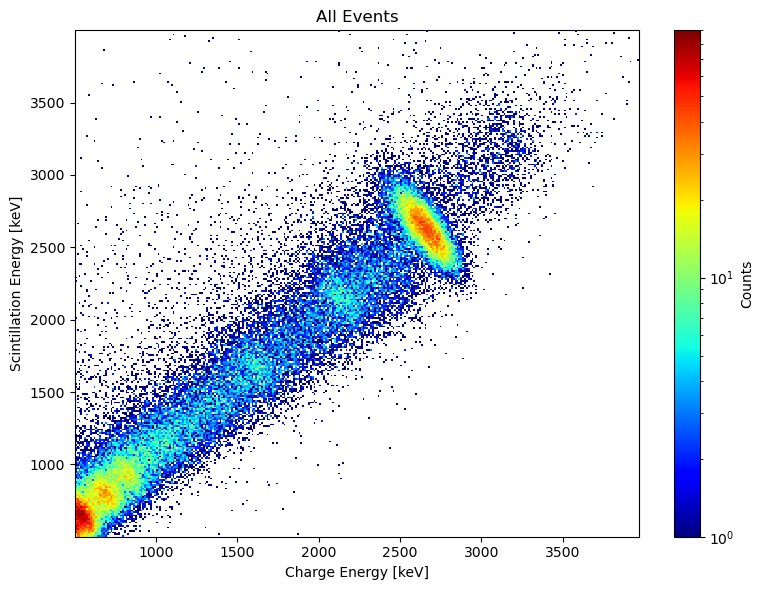}
    \vspace{0.3em}
    
\end{minipage}
\hfill
\begin{minipage}[t]{0.48\textwidth}
    \centering
    \includegraphics[width=\linewidth]{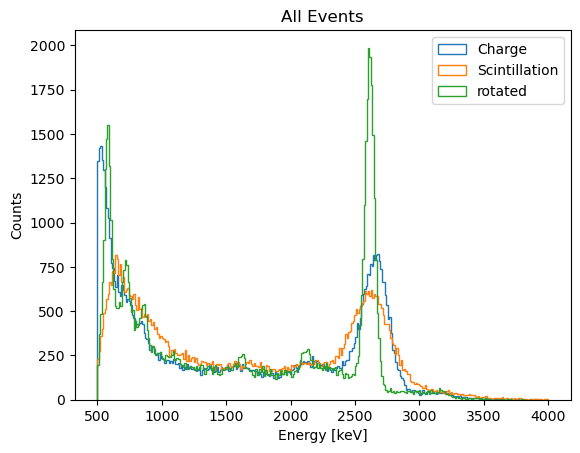}
    \vspace{0.3em}
    
\end{minipage}

\caption{Left: Two-dimensional histogram of scintillation energy versus charge energy for Thorium source calibration run 8970, illustrating the anticorrelation between the two signals. Right: Corresponding energy spectra reconstructed using charge energy, scintillation energy, and their optimized linear combination (rotated energy).
}
\label{fig:Th-source}
\end{figure*}

The open-source dataset is stored in the HDF5 (\texttt{.hdf5}) format, which is widely used for large scientific datasets and machine-learning workflows. The HDF5 structure naturally supports hierarchical storage of waveform arrays, reconstructed quantities, and metadata within a single portable file format. The entire dataset is publicly hosted on Zenodo~\cite{zenodo}, an open-access scientific data repository widely used for long-term archival and citation of research datasets. The total size of the open-source dataset is approximately 70 GB, including waveform-level detector data and reconstructed event quantities. To access the dataset, please use the following \url{https://doi.org/10.5281/zenodo.22155860}. Each run is stored in a separate HDF5 file named \texttt{run\_number.h5}, where \texttt{run\_number} denotes the EXO-200 run number. An example Jupyter notebook, \texttt{load\_h5.ipynb}, is provided to demonstrate how to load the EXO-200 HDF5 dataset. The script gives examples of plotting individual channel waveforms, the calibration source energy spectra, and event locations. Figure ~\ref{fig:Th-source} shows the energy spectra from one of the source runs in the open dataset, plotted using the example notebook. 

\section{Disclaimer}\label{sec:disclaimer}
The EXO-200 collaboration has authorized the public release of this dataset. The dataset may be used in accordance with the license specified in the Zenodo record. Individuals or collaborations are allowed to publish papers based on this dataset without including the EXO-200 collaborators as authors.  The EXO-200 collaboration retains ownership and stewardship of this dataset and associated documentation. Users of this dataset are kindly requested to cite the primary EXO-200 publications associated with the open-source dataset, including Ref.~\cite{Anton_2019}, as well as the dataset DOI 10.5281/zenodo.22155860,  and this ArXiv posting.

\begin{acknowledgments}
This work was supported by grants from the NSF
(PHY-2111213 and PHY-2514767) as well as from the
DOE (DE-SC0020509). EXO-200 was supported by the DOE (DE-SC0017970, DE-FG02-01ER41166, DESC0021383, DE-SC0014517) and NSF in the U.S., the NSERC in Canada (CRC-2019-00399, SAPPJ-2017-00029), the SNF in Switzerland, the IBS in Korea, the RFBR (18-02-00550) in Russia, the DFG in Germany, and CAS and ISTCP in China. The EXO-200 data analysis and simulation use resources of the SLAC Shared Scientific Data Facility (S3DF). We gratefully acknowledge the KARMEN collaboration for supplying the cosmic-ray veto detectors, as well as the WIPP for their hospitality.
\end{acknowledgments}

\bibliography{BDT_unidoc}

@article{EXO-0nbb-12,
  title = {Search for Neutrinoless Double-Beta Decay in $^{136}\mathrm{Xe}$ with {EXO}-200},
  author = {Auger, M. and others},
  collaboration = {EXO-200 Collaboration},
  journal = {Phys. Rev. Lett.},
  volume = {109},
  issue = {3},
  pages = {032505},
  numpages = {6},
  year = {2012},
  month = {Jul},
  publisher = {American Physical Society},
  doi = {10.1103/PhysRevLett.109.032505},
  url = {https://link.aps.org/doi/10.1103/PhysRevLett.109.032505}
}

@article{EXO-200-onbb-14,
    author = "Albert, J. B. and others",
    collaboration = "EXO-200 Collaboration",
    title = "{Search for Majorana neutrinos with the first two years of {EXO}-200 data}",
    eprint = "1402.6956",
    archivePrefix = "arXiv",
    primaryClass = "nucl-ex",
    doi = "10.1038/nature13432",
    journal = "Nature",
    volume = "510",
    pages = "229--234",
    year = "2014"
}

@article{Albert_2018,
   title="{Search for Neutrinoless Double-Beta Decay with the Upgraded {EXO}-200 Detector}",
   volume={120},
   ISSN={1079-7114},
   url={http://dx.doi.org/10.1103/PhysRevLett.120.072701},
   DOI={10.1103/physrevlett.120.072701},
   number={7},
   journal={Physical Review Letters},
   publisher={American Physical Society (APS)},
   author={Albert, J. B. and others},
    collaboration = "EXO-200 Collaboration",
   year={2018},
   month=Feb }

@phdthesis{shaolei_thesis,
    author = "Li, Shaolei",
    title = {Search for Neutrinoless Double Beta Decay with EXO-200 and the Application of Deep Learning to Detector Simulation},
    school = {University of Illinois at Urbana-Champaign},
    year = 2021,
    url = {https://hdl.handle.net/2142/115319}
}

@misc{nelson,
  title = {MEMORANDUM FOR THE HEADS OF EXECUTIVE DEPARTMENTS AND AGENCIES},
  author = {Dr. Alondra Nelson},
  collaboration = {Office of Science and Technology Policy (OSTP)},
  year = {2022},
  month = {August},
  publisher = {Executive Office of the President},
  note = {\url{https://www.whitehouse.gov/wp-content/uploads/2022/08/08-2022-OSTP-Public-Access-Memo.pdf}}
}

@misc{cerati2023microboonepublicdatasets,
      title={{MicroBooNE} Public Data Sets: a Collaborative Tool for {LArTPC} Software Development}, 
      author={Giuseppe Cerati},
      year={2023},
      eprint={2309.15362},
      archivePrefix={arXiv},
      primaryClass={hep-ex},
      url={https://arxiv.org/abs/2309.15362}, 
}

@article{Bukhari2024IceCube,
  author  = {Bukhari, Habib and Chakraborty, Dipam and Eller, Philipp
             and Ito, Takuya and Shugaev, Maxim V. and {\O}rs{\o}e, Rasmus},
  title   = {IceCube -- Neutrinos in Deep Ice: The Top 3 Solutions
             from the Public Kaggle Competition},
  journal = {Eur. Phys. J. C},
  volume  = {84},
  pages   = {646},
  year    = {2024},
  doi     = {10.1140/epjc/s10052-024-12977-2},
  eprint  = {2310.15674},
  archivePrefix = {arXiv}
}

@misc{COHERENT,
    author = {Akimov, D. and others},
    collaboration = {COHERENT Collaboration},
    title = {{COHERENT Collaboration data release from the first detection of coherent elastic neutrino-nucleus scattering on argon}},
    eprint = {2006.12659},
    archivePrefix = {arXiv},
    journal = "",
    primaryClass = {nucl-ex},
    doi = {10.5281/zenodo.3903810},
    month = {6},
    year = {2020}
}

@misc{MJD-datarelease,
      title={{Majorana Demonstrator data release for AI/ML applications}}, 
      author={I. J. Arnquist and others},
      collaboration = {Majorana Collaboration},
      year={2023},
      eprint={2308.10856},
      archivePrefix={arXiv},
      primaryClass={cs.LG},
      url={https://arxiv.org/abs/2308.10856}, 
    doi = {10.5281/zenodo.8257027},
}

@article{Ackerman_2011,
   title={Observation of Two-Neutrino Double-Beta Decay in {Xe}-136 with the {EXO}-200 Detector},
   volume={107},
   ISSN={1079-7114},
   url={http://dx.doi.org/10.1103/PhysRevLett.107.212501},
   DOI={10.1103/physrevlett.107.212501},
   number={21},
   journal={Physical Review Letters},
   publisher={American Physical Society (APS)},
   author={Ackerman, N. and others},
    collaboration = {EXO-200 Collaboration},
   year={2011},
   month=Nov }

@article{EXO-Detector_2012,
   title={The {EXO}-200 detector, part {I}: detector design and construction},
   volume={7},
   ISSN={1748-0221},
   url={http://dx.doi.org/10.1088/1748-0221/7/05/P05010},
   DOI={10.1088/1748-0221/7/05/p05010},
   number={05},
   journal={Journal of Instrumentation},
   publisher={IOP Publishing},
   author={Auger, M and others},
   collaboration = {EXO-200 Collaboration},
   year={2012},
   month=May, pages={P05010–P05010} }

@article{EXO-detector_2022,
   title={The {EXO-200} detector, part {II}: auxiliary systems},
   volume={17},
   ISSN={1748-0221},
   url={http://dx.doi.org/10.1088/1748-0221/17/02/P02015},
   DOI={10.1088/1748-0221/17/02/p02015},
   number={02},
   journal={Journal of Instrumentation},
   publisher={IOP Publishing},
   author={Ackerman, N. and others},
   collaboration = {EXO-200 Collaboration},
   year={2022},
   month=Feb, pages={P02015} }

@article{Albert_2014,
   title={An improved measurement of the 2$\nu\beta\beta$ half-life of {Xe}-136 with {EXO-200}},
   volume={89},
   ISSN={1089-490X},
   url={http://dx.doi.org/10.1103/PhysRevC.89.015502},
   DOI={10.1103/physrevc.89.015502},
   number={1},
   journal={Physical Review C},
   publisher={American Physical Society (APS)},
   author={Albert, J. B. and others},
   year={2014},
   collaboration = {EXO-200 Collaboration},
   month=Jan }

@article{Anton_2019,
   title={Search for Neutrinoless Double-Beta Decay with the Complete {EXO-200} Dataset},
   volume={123},
   ISSN={1079-7114},
   url={http://dx.doi.org/10.1103/PhysRevLett.123.161802},
   DOI={10.1103/physrevlett.123.161802},
   number={16},
   journal={Physical Review Letters},
   publisher={American Physical Society (APS)},
   author={Anton, G. and others },
   collaboration = {EXO-200 Collaboration},
   year={2019},
   month=Oct }

@article{Davis_2016,
   title={An optimal energy estimator to reduce correlated noise for the {EXO}-200 light readout},
   volume={11},
   ISSN={1748-0221},
   url={http://dx.doi.org/10.1088/1748-0221/11/07/P07015},
   DOI={10.1088/1748-0221/11/07/p07015},
   number={07},
   journal={Journal of Instrumentation},
   publisher={IOP Publishing},
   author={Davis, C.G. and others},
   collaboration = {EXO-200 Collaboration},
   year={2016},
   month=July, pages={P07015–P07015} }

@article{Delaquis2018,
  title = {Deep neural networks for energy and position reconstruction in {EXO}-200},
  volume = {13},
  ISSN = {1748-0221},
  url = {http://dx.doi.org/10.1088/1748-0221/13/08/P08023},
  DOI = {10.1088/1748-0221/13/08/p08023},
  number = {08},
  journal = {Journal of Instrumentation},
  publisher = {IOP Publishing},
  author = {Delaquis,  S. and others},
  collaboration = {EXO-200 Collaboration},
  year = {2018},
  month = Aug,
  pages = {P08023–P08023}
}

@article{Neilson_2009,
   title={Characterization of large area {APDs} for the {EXO}-200 detector},
   volume={608},
   ISSN={0168-9002},
   url={http://dx.doi.org/10.1016/j.nima.2009.06.029},
   DOI={10.1016/j.nima.2009.06.029},
   number={1},
   journal={Nuclear Instruments and Methods in Physics Research Section A: Accelerators, Spectrometers, Detectors and Associated Equipment},
   publisher={Elsevier BV},
   author={Neilson, R. and others},
     collaboration = {EXO-200 Collaboration},
   year={2009},
   month=Sept, pages={68–75} }

@misc{Zenodo,
  author    = {{European Organization for Nuclear Research} and {OpenAIRE}},
  title     = {Zenodo},
  publisher = {CERN},
  year      = {2013},
  doi       = {10.25495/7GXK-RD71}
}

\end{document}